\documentclass[11pt]{article}

\usepackage{times}
\usepackage{pdfpages}
\usepackage{setspace}
\usepackage{sectsty}
\usepackage{enumitem}

\sectionfont{\fontsize{11}{15}\selectfont}
\subsectionfont{\fontsize{11}{15}\selectfont}

\usepackage{wrapfig}
\usepackage{lipsum}
\usepackage{graphicx}

\usepackage{array}
\usepackage{subfigure}
\usepackage{booktabs}
\usepackage{multicol}
\usepackage{multirow}
\usepackage{threeparttable}
\usepackage[small, bf]{caption}
\usepackage[tableposition=top]{caption}
\usepackage{booktabs}
\usepackage{siunitx}

\usepackage{pgfgantt}

\usepackage{wrapfig}
\usepackage{times}
\usepackage{url}
\usepackage{soul}
\usepackage{color}
\begingroup
\makeatletter
\g@addto@macro{\UrlSpecials}{%
  \endlinechar=13 \catcode\endlinechar=12
  \do\%{\Url@percent}\do\^^M{\break}}
 \gdef\Url@percent{\@ifnextchar^^M{\@gobble}{\mathbin{\mathchar`\%}}}%
\endgroup %
\usepackage{array}
\usepackage{subfigure}

\usepackage{array}
\newcolumntype{L}[1]{>{\raggedright\let\newline\\\arraybackslash\hspace{0pt}}m{#1}}

\usepackage[headheight=14pt,tmargin=1in,bmargin=1in,lmargin=1in,rmargin=1in]{geometry}

\usepackage{fancyhdr}
\usepackage{datetime}
\usepackage{indentfirst}
\usepackage[sort&compress, square, comma, numbers]{natbib}

\usepackage[compact]{titlesec}

\usepackage{titlesec}
\titlespacing*{\section}{0pt}{0.1\baselineskip}{0.1\baselineskip}
\titlespacing*{\subsection}{0pt}{0.2\baselineskip}{0.1\baselineskip}
\titlespacing*{\subsubsection}{0pt}{0.2\baselineskip}{0.1\baselineskip}
\titlespacing*{\paragraph}{0pt}{0.2\baselineskip}{0.2\baselineskip}

\usepackage{blindtext}

\usepackage{threeparttable}
\usepackage{soul}
\usepackage{booktabs}
\usepackage{multirow}
\usepackage{caption}
\usepackage{graphicx}
\usepackage{graphbox}
\usepackage{paralist}
\usepackage[section]{placeins}
\begin{document}

\thispagestyle{empty}
\begin{center}
{\bf \Large Discovering the Traces of Disinformation on Instagram in the Internet Archive}\\

\vspace{5mm}
Haley Bragg, haley.bragg.19@cnu.edu\\
Department of Computer Science, Christopher Newport University\\
\hfill \break
Dr. Michele Weigle, mweigle@cs.odu.edu\\
Department of Computer Science, Old Dominion University\\
\end{center}

\begin{abstract}
  Disinformation, which is fabricated, misleading content spread with the intent to deceive others, is accumulating substantial engagements and reaching a vast audience on Instagram. However, the temporary nature of the platform and the security guidelines that remove malicious content make studying this disinformation a challenge. The only way to access removed content and banned accounts that are no longer on the live web is by searching the web archives. In this study, we set out to quantify the replayability and quality of past captures of Instagram account pages, specifically focusing on a group of anti-vaxx content creators known as the Disinformation Dozen. We found that the number of mementos listed for these users' account pages on the Internet Archive's Wayback Machine can be misleading, because a majority of the mementos are actually redirections to the Instagram login page, and of the remaining replayable mementos, many are missing post images. In fact, 96.13\% of mementos from the Disinformation Dozen accounts redirect to the login page, and only 27.16\% of the remaining replayable mementos contain every post image. Combined, these results reveal that merely 1.05\% of mementos for the Disinformation Dozen accounts are replayable with complete post images. Furthermore, we found that the percentage of replayable mementos is decreasing over time, with a particular lack of replayable mementos for the years 2021 and 2022.

\end{abstract}
\section{Introduction}

\begin{figure}[h!]
\centering
\includegraphics[width=1.0\linewidth]{ 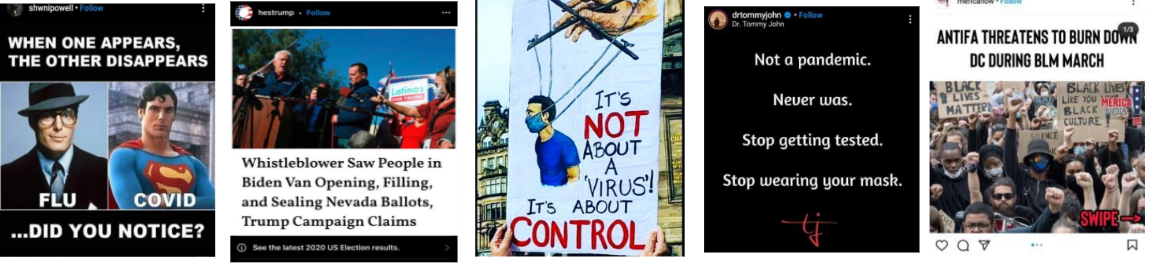}
\vspace*{-7mm}
\caption{Examples of Disinformation on Instagram. Images from \cite{CCDH2021disinformation} \cite{CCDH2021malgorithm}}
\label{fig:disinfo_examples}
\end{figure}

The success of Instagram as a photo-centric social media platform has revolutionized the speed and ease at which information can be consumed online. With easily digestible memes, striking political art, and aesthetic infographics, Instagram epitomizes the phrase ``a picture is worth a thousand words." However, as with most technological advancements, people with malicious intentions can also use the space for their deceitful agendas. For example, certain wrongdoers are taking advantage of the 1.21 billion monthly active users \cite{dixon2020instagram} to propagate disinformation, which can be defined as ``information that is false and deliberately created to harm a person, social group, organization or country" \cite{ireton2018journalism}. Figure \ref{fig:disinfo_examples} portrays examples of this disinformation, from COVID-19 memes, to conspiracy theory political art, to inciteful, misleading headlines about the 2020 election and BLM protests.

This issue is persistent on other social media platforms like Twitter and Facebook, of course, but comparatively, it is both overlooked and understudied on Instagram. For example, the Internet Research Agency, a Russian group that attempted to influence American political discourse about the 2016 election with divisive social media campaigns, utilized all three platforms in their efforts. However, when the scandal went before the US Senate, only Twitter and Facebook were thoroughly discussed, even though the Instagram disinformation received more engagements than the Facebook and Twitter engagements combined (187 million for Instagram, 77 million for Facebook, and 73 million for Twitter) \cite{diresta2019tactics}. As for why Instagram is understudied, multiple factors play a role. For one, Instagram content has a more temporary nature; whereas posts are shared permanently via retweeting and quote tweeting on Twitter, posts are shared via stories that disappear after 24 hours on Instagram, creating a much smaller window for storing the content. Additionally, Instagram allows users to edit their captions, tags, and locations, whereas Twitter has no current edit feature, solidifying the nonpermanent nature of the photo-sharing app. Another hardship for studying Instagram is the absence of a feature to share links. Mapping link sharing is a widely-employed method for studying the spread of information, and leading disinformation scholar Kate Starbird utilized this method in her study of alternative media echo chambers on Twitter \cite{starbird2018ecosystem}. Since Instagram does not hyperlink URLs in post captions, it is more difficult for researchers to establish these maps of networks and connections. Finally, researchers have found that Instagram is harder to archive than its counterparts \cite{jayanetti_2020}. 

This last issue is particularly relevant to the study of disinformation on Instagram. Due to the nature of the malicious content, a lot of posts spreading disinformation are caught by fact-checking software and deleted. Similarly, accounts dedicated to disinformation are banned once they violate guidelines repeatedly. For example, the Center for Countering Digital Hate released a report in 2021 on a group of content creators known as the ``Disinformation Dozen," \cite{CCDH2021disinformation} who were responsible for 65\% of anti-vaxx content online (full list of members in Table \ref{table:list_of_members}). Since the release of the center's report, Instagram has taken measures to address the group \cite{spokesperson2021banned}, and 10 out of 12 of the disinformation actors have been banned from Instagram, meaning none of their posts are available on the live web anymore. However, web archiving keeps this content from being lost forever, storing captures of the pages as they once appeared on the live web. In this study\footnote{This work was performed during the NSF REU Site in Disinformation Detection and Analytics held at Old Dominion University, Summer 2022.}, we quantify how well these archived captures, known as mementos, represent the content that was posted by the Disinformation Dozen.

\begin{table}[!h]
\centering
\begin{tabular}{|c|}
\hline
\textbf{Disinformation Dozen Members} \\
\hline
Joseph Mercola \\
Robert F. Kennedy, Jr. \\
Ty and Charlene Bollinger \\
Sherri Tenpenny \\
Rizza Islam \\
Rashid Buttar \\
Erin Elizabeth \\
Sayer Ji \\
Kelly Brogan \\
Christiane Northrup \\
Ben Tapper \\
Kevin Jenkins \\
\hline
\end{tabular}
\vspace{2mm}
\caption{Disinformation Dozen Members}
\label{table:list_of_members}
\end{table}

\section{Related Work}
To quantify archivists' suspicions regarding Instagram's underperformance in the web archiving sphere, Himarsha Jayanetti, PhD student at Old Dominion University, performed the case study ``How well is Instagram archived?" in November of 2020 \cite{jayanetti_2020}. She chose to focus on Katy Perry's Instagram account because Perry was the 20th most popular person on Instagram at the time of the study and was posting the same content on all three leading social media platforms (Instagram, Facebook, and Twitter). This uniformity allowed Jayanetti to directly compare the number of mementos of Perry's different social media account pages to determine whether Instagram had fewer captures than Facebook and Twitter. As seen below in Table \ref{table:katy_perry_captures}, the results indicate that Instagram had the smallest number of mementos, quantitatively solidifying presumptions that Instagram is not as well-archived as Facebook and Twitter. Furthermore, Himarsha reported that only 31.67\% of Perry's posts were archived in public web archives, and noted that she observed a considerable amount of archiving and replayability issues for mementos of Instagram pages. 

\begin{center}
\begin{table}[!h]
\centering
\begin{tabular}{|c|c|c|c|}
\hline
& Instagram & Facebook & Twitter \\
\hline
Internet Archive & 1803 & 2032 & 4025 \\
Archive-It & 108 & 1234 & 1185 \\
archive.today & 7 & 0 & 2 \\
Library of Congress & 4 & 1 & 0 \\
UK Web Archive & 0 & 16 & 58  \\
Australian Web Archive & 0 & 37 & 2 \\
Portuguese Web Archive & 2 & 0 & 1 \\
\hline
Total & 1924 & 3320 & 5273 \\
\hline
\end{tabular}
\vspace{3mm}
\caption{Captures of Katy Perry's social media account pages across web archives. Table reproduced from \cite{jayanetti_2020}}
\label{table:katy_perry_captures}
\end{table}
\end{center}
\vspace{-10mm}

\section{Methodology}
\subsection{Comparing Number of Mementos Across Social Media Platforms}
To gain an initial understanding of how well the Disinformation Dozen accounts are archived, we repeated Jayanetti's analysis to quantify how many mementos of a given account are available across different web archives.  The Instagram, Facebook, and Twitter handles that we used for this analysis are available in the Appendix in Table \ref{table:social_media_handles}. This analysis utilized MemGator \cite{jcdl-2016:alam:memgator}, which aggregates mementos for a certain URL from 16 different web archives, including the Internet Archive, the Library of Congress, and archive.today. As predicted, several accounts paralleled Himarsha's finding that Instagram has the lowest number of mementos when compared with Facebook and Twitter, such as those of Joseph Mercola, Sayer Ji, Kelly Brogan, and Christiane Northrup. However, we also identified four accounts for which Instagram had the highest number of mementos when compared with Facebook and Twitter: Robert F. Kennedy Jr., Sherri Tenpenny, Rizza Islam, and Ben Tapper (the data tables for all Disinformation Dozen accounts are available in the Appendix in Tables \ref{table:joseph_mercola_captures}-\ref{table:ben_tapper_captures}). This finding may be explained by the fact that the members of the Disinformation Dozen do not share the same content across all three platforms. Instead, these malicious content creators may show a preference for one particular social media site, in this case Instagram, which would result in more traffic on their account page for that platform, and consequently, relatively more mementos. 

\subsection{Determining How Well Instagram Mementos Replay}

After establishing this baseline number of mementos for each Instagram account of the Disinformation Dozen, our main goal was to evaluate how many of these mementos were actually ``good" or ``bad." This goal is similar to evaluating a memento's damage, as investigated in a study by Brunelle et al. \cite{6970187}. The image below shows a ``good" capture of an Instagram account page, with nine visible post images intact.

\begin{figure}[h!]
\centering
\includegraphics[width=0.3\linewidth]{ 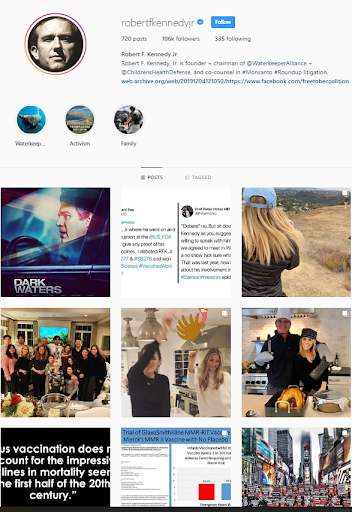}
\vspace*{-3mm}
\caption{A Memento of an Instagram Account Page with Nine Post Images}
\label{fig:robert_kennedy_memento}
\end{figure}

We set out to answer the following questions: Is every memento a full capture of a past Instagram page, or are there replayability issues? Are there differing qualities of mementos in terms of missing resources? To answer these questions, we have chosen to use the banned Instagram account of Robert F. Kennedy Jr. \cite{chappell2021banned} as an example.  With 598 mementos captured by the Internet Archive, Kennedy is the member of the Disinformation Dozen with the most Instagram mementos, and therefore most favorable for illustrative purposes.

\subsubsection{Visualizing Mementos with the Wayback Calendar View}
\begin{figure}[h!]
\centering
\includegraphics[width=0.80\linewidth]{ 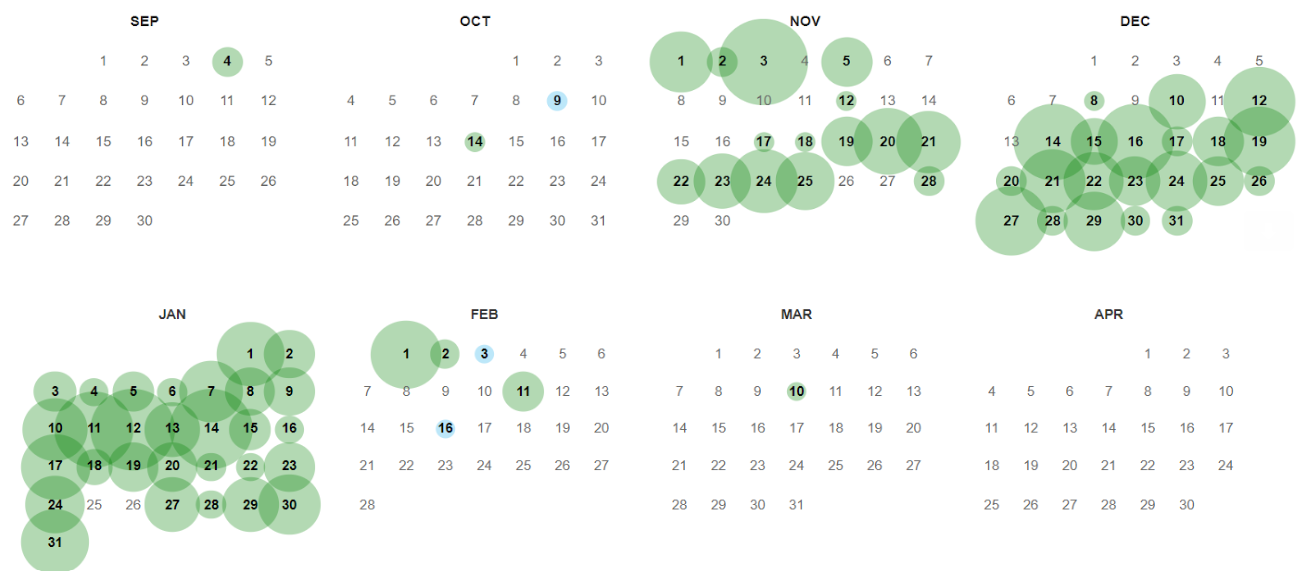}
\vspace*{-3mm}
\caption{Wayback Machine Calendar View 2020-2021 for Kennedy}
\label{fig:calendar_view}
\end{figure}

The Internet Archive's Wayback Machine includes a calendar view that maps each time a given URL was crawled by the Wayback Machine as a circle around the corresponding date it was crawled. The larger the circle, the more mementos available for that day (see November 3 in Figure \ref{fig:calendar_view}, having 32 mementos). The calendar view for Robert F. Kennedy Jr.'s Instagram page shows the months leading up to his removal from Instagram in February 2021; the large quantity and size of the circles seemingly implicate that this URL is well-archived. However, the color of the circles is a significant indicator of the replayability of these archived captures. A green circle designates that a majority of the capture attempts received a HTTP 3xx response code, which indicates a redirect, whereas a blue circle designates that a majority of the capture attempts received a HTTP 2xx response code, which indicates that the request succeeded.

\begin{figure}[h!]
\centering
\begin{minipage}[b]{0.4\linewidth}
\includegraphics[width=\textwidth]{ 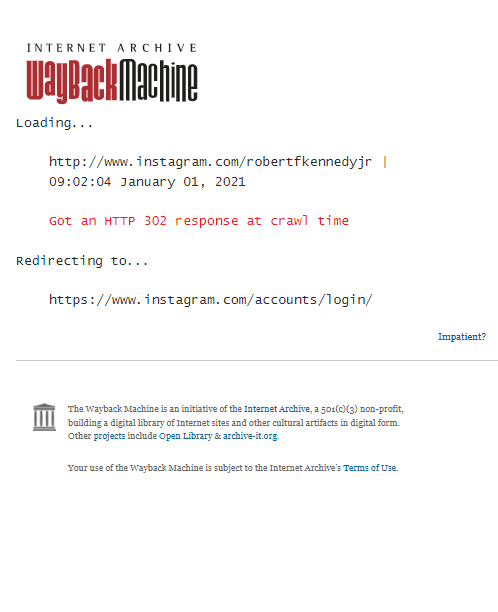}
\caption{Wayback Machine Redirect Message}
\label{fig:redirect_message}
\end{minipage}
\hspace{0.5cm}
\begin{minipage}[b]{0.35\linewidth}
\centering
\includegraphics[width=\textwidth]{ 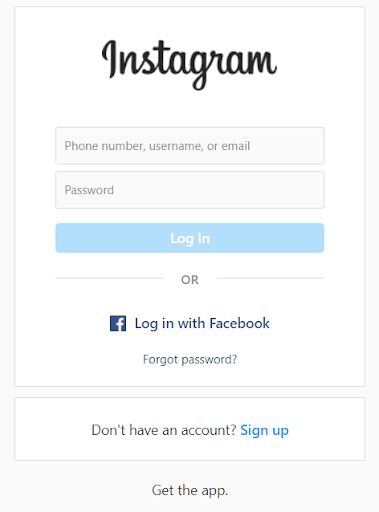}
\caption{Instagram Login Page}
\label{fig:login_page}
\end{minipage}
\end{figure}

With regard to Instagram specifically, the large quantity of green circles seen above are caused by redirects to the Instagram login page. Instagram has strict security measures in place against users who are not logged in, so any non-logged in user browsing the platform is eventually redirected to the login page. In this case, the Wayback crawler is acting as a non-logged in user and is diverted accordingly. Attempting to follow the link in a green circle results in the message shown in Figure \ref{fig:redirect_message}, and a consequent redirection to the login screen (Figure \ref{fig:login_page}). The login page is clearly not a capture of Robert F. Kennedy Jr.'s Instagram account and provides no information about the banned content he was posting. In contrast, following a successful link in a blue circle allows Wayback Machine users to see a legitimate capture of the past, as if they were viewing Kennedy's account at the time. Figure \ref{fig:robert_kennedy_memento} exemplifies one such successful memento with all post images intact. 

\vspace{0.5cm}
\subsubsection{Quantifying Redirects, Successes, and Warc/Revisits}
\begin{figure}[h!]
\centering
\includegraphics[width=0.9\linewidth]{ 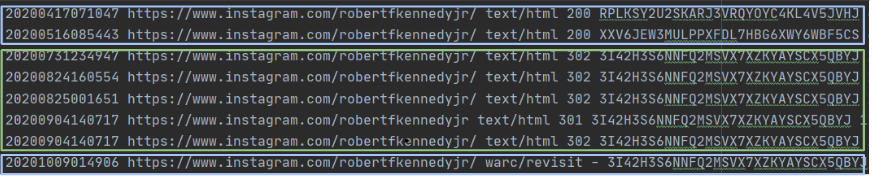}
\vspace*{-3mm}
\caption{Sample CDX Data for Robert F. Kennedy Jr. }
\label{fig:cdx_data}
\end{figure}

After establishing the difference between the ``green" mementos that redirect and the ``blue" mementos that replay, a natural next step was quantifying the number of mementos that belonged to each category. Specifically, we wanted to determine how many of the 598 archived captures for Robert F. Kennedy's Instagram account page redirected to the login screen, versus how many were successful captures of his banned account. Even though one could technically manually classify each memento, that approach would be time-consuming and scale poorly for larger data sets. Instead, we automated the process using the Internet Archive's CDX API\footnote{https://github.com/internetarchive/wayback/tree/master/wayback-cdx-server}, which returns the raw CDX format shown in Figure \ref{fig:cdx_data}. It is important to note that the calendar page is actually populated by the CDX information, so the same data is being evaluated. In the CDX format, each line corresponds with a different memento and includes the HTTP response code as the fifth space-separated field. We extracted this status code, classified the mementos with 200 response codes as ``successes," and the mementos with 301 or 302 response codes as ``redirects." The results are shown in Table \ref{table:redirects_vs_successes}. 
\pagebreak

It is worth highlighting that some mementos had a response code of \texttt{-} and a MIME type of \texttt{warc/revisit} (see bottom line of Figure \ref{fig:cdx_data}). Even though these mementos appear as blue circles on the Wayback Machine calendar view, they are actually duplicates of material already archived, including both redirects to the login page and replayable mementos. One can identify revisits to redirects by either using a curl command and grepping for ``Got an HTTP 302 response at crawl time$\vert$Login" or checking the sixth field of the CDX object for the hash that characterizes the login page (\texttt{3I42H3S6NNFQ2MSVX7XZKYAYSCX5QBYJ}).
\vspace{-5mm}

\begin{table}[!h]
\centering
\vspace*{0.75 cm}
\begin{tabular}{|c|c|c|}
\hline
& Number of Mementos & Percentage of Total Mementos \\
\hline
Identified Redirects & 574 & 95.99\% \\
Warc/Revisits to Redirects & 18 & 3.01\% \\
\hline
Total Redirects & 592 & 99.0\% \\
\hline
Identified Successes & 6 & 1.0\% \\
Warc/Revisits to Successes & 0 & 0.0\% \\
\hline
Total Successes & 6 & 1.0\% \\
\hline
Total Mementos & 598 & 100.0\% \\
\hline
\end{tabular}
\caption{Redirects vs. Successes for instagram.com/RobertFKennedyJr}
\label{table:redirects_vs_successes}
\end{table}
\vspace{-3mm}

As seen in Table \ref{table:redirects_vs_successes}, we found that 574 out of the 598 mementos for Kennedy's Instagram page were the ``green" redirects to the login screen. That is, 95.99\% of the supposed archived captures of his Instagram page are not captures of his Instagram page at all. Furthermore, of the 24 remaining mementos that appeared blue on the Wayback Machine's calendar view, 18 were classified as warc/revisits to redirects, meaning the number of Kennedy's mementos that redirected to the login screen is actually 592. These results reveal that only 6 of Kennedy's mementos, which is 1.0\% of his total mementos, are replayable captures. Similar tables for all members of the Disinformation Dozen are available in the Appendix (Tables \ref{table:joseph_mercola_redirects}-\ref{table:ben_tapper_redirects}).
\vspace{2mm}

\subsubsection{Quality of the Six Replayable Mementos}
\label{section:post_images}
\begin{figure}[h!]
\begin{minipage}[b]{0.293\linewidth}
\includegraphics[width=\textwidth]{ 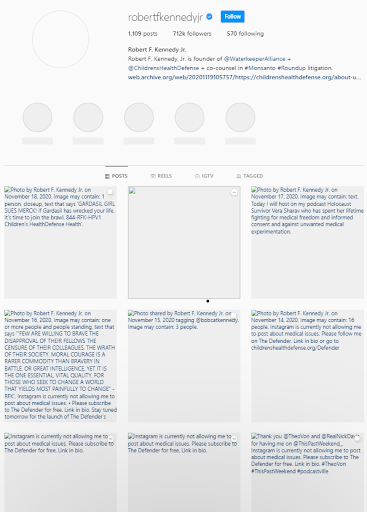}
\caption{Memento with No Loaded Post Images}
\label{fig:no_loaded_post_images}
\end{minipage}
\hspace{0.5cm}
\begin{minipage}[b]{0.308\linewidth}
\centering
\includegraphics[width=\textwidth]{ 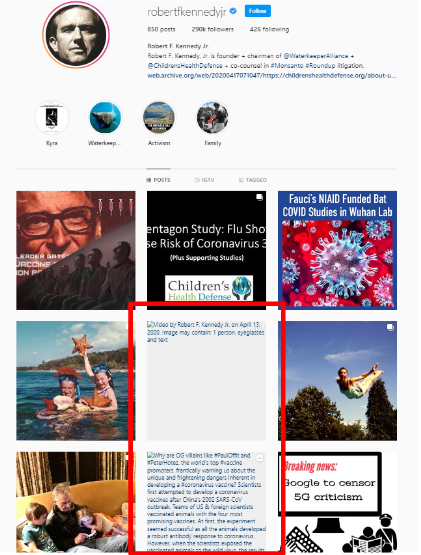}
\caption{Memento with\\ Incomplete Loaded Post Images}
\label{fig:some_loaded_post_images}
\end{minipage}
\hspace{0.5cm}
\begin{minipage}[b]{0.3\linewidth}
\centering
\includegraphics[width=\textwidth]{ 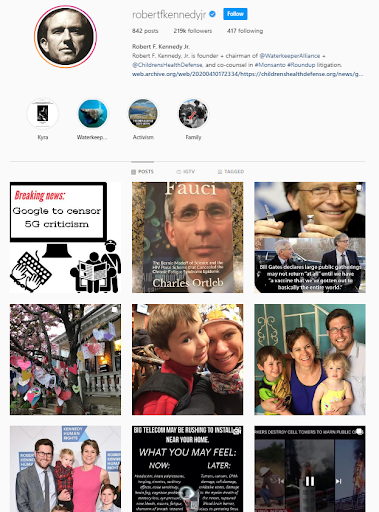}
\caption{Memento with Complete Loaded Post Images}
\label{fig:complete_loaded_post_images}
\end{minipage}
\end{figure}

 Even though we identified the number of mementos that do not redirect to the login page, we still must ask the following question: Does replayability guarantee a complete capture, fully representative of the content a user was posting? Since Instagram is centered around sharing images, quantifying the number of post images that are present in the memento is a relevant measure of this completeness. For example, three of the six replayable mementos for Kennedy's account have zero post images available, as seen in Figure \ref{fig:no_loaded_post_images}. Even though alternative text is available for some posts, the pictures| which are the very essence of this photo-sharing platform| are missing. Similarly, one of the six replayable mementos had only a partial number of photo images available (Figure \ref{fig:some_loaded_post_images}). While the quality of this memento is relatively better than the three with no loaded post images, it is still not a complete capture of the past. This brings us to our culminating observation regarding the number of mementos for Robert F. Kennedy's Instagram page: only two mementos out of the original 598 are complete, replayable mementos. These two mementos, with every post image intact (see Figure \ref{fig:complete_loaded_post_images}), are the only complete captures of Robert F. Kennedy's Instagram account page as it once appeared on the live web.

\vspace*{0.1 cm}

\subsection{Programmatically Counting the Number of Post Images that Replay}
\label{section:counting_post_images}
\begin{figure}[h!]
\centering
\includegraphics[width=0.85\linewidth]{ 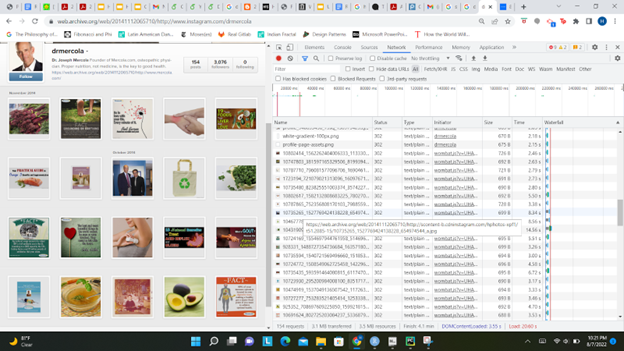}
\vspace*{-3mm}
\caption{Locating Post Image Requests in Network Activity Tab}
\label{fig:network_activity}
\end{figure}

As described in Section \ref{section:post_images}, the number of post images that replay for a given memento is indicative of that memento's quality. However, counting the number of loadable pictures manually would be at best a tedious task, and for larger data sets, an infeasible one. This infeasibility  motivated our development of a Python script\footnote{https://github.com/haleybragg/DisinformationDozen} to automate the process. The first step was inspecting the Network Activity tab in Google Chrome to identify which resource requests corresponded with the post images. As pictured in Figure \ref{fig:network_activity}, the post image resources each seemed to have a unique, identifying sequence of numbers and underscores, and in this particular case from 2014, each ended in ``\_a.jpg." We found that this ending was not consistent, however, changing to ``\_s.jpg" in early 2015, ``\_n.jpg" in late 2015, then ``\_n.jpg" moved altogether to the middle of the resource URL in 2016. 

Once we determined each possible case for the name of the post image resources, we used the Selenium Wire\footnote{https://pypi.org/project/selenium-wire/} Python package to programmatically access these network requests made by the Chrome browser. We iterated through each request for each replayable memento, using regular expressions to extract the identifying sequence of numbers for each post image. It must be understood that an image ID can occur in multiple requests per memento because the response for a particular image request can be a redirection to a nearby memento. In terms of HTTP status codes, this occurrence could result in a string of ``302" response codes until the post image request is finally successful or unsuccessful. For this reason, we allowed for multiple status codes to be associated with a particular photo ID in a dictionary, and if ``200" was present in the list of status codes, we knew that image would display correctly in the archived capture. The number of post images that replayed successfully and the total number of post images for each memento were written to a .csv file, which was then uploaded to R for aggregation and visualization.

\begin{figure}[!h]
\centering
\includegraphics[width=0.9\linewidth]{ 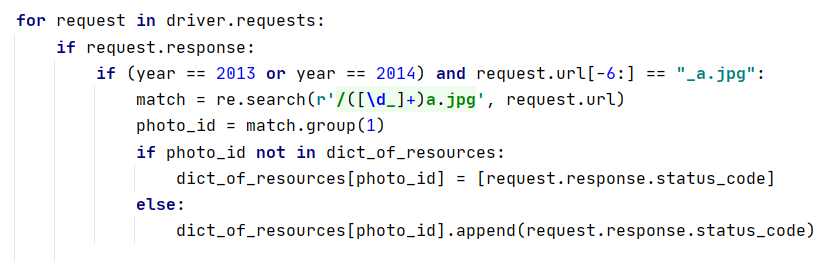}
\vspace*{-3mm}
\caption{Sample Code for Storing Status Codes Per Image}
\label{fig:image_code}
\end{figure}

\section{Results}
Not only are these replayability and quality issues present for the mementos of Robert F. Kennedy Jr.'s Instagram account page, but also they constitute broader trends for Instagram archivability over the years. To examine these trends, we aggregated data for both the collection of Disinformation Dozen Instagram accounts and a collection of Health Authorities listed in the Center for Countering Digital Hate's Malgorithm report\cite{CCDH2021malgorithm}, which includes organizations like the Center for Disease Control and the World Health Organization. A full list of the Health Authorities' Instagram accounts is available in the Appendix in Table \ref{table:instagram_handles}. Each of the following graphs that illustrate such trends were generated using the ggplot2 package for R. 

\subsection{How Many Mementos Redirect for the Disinformation Dozen and the Health Authorities?}
\begin{figure}[!h]
\centering
\includegraphics[width=0.8\linewidth]{ 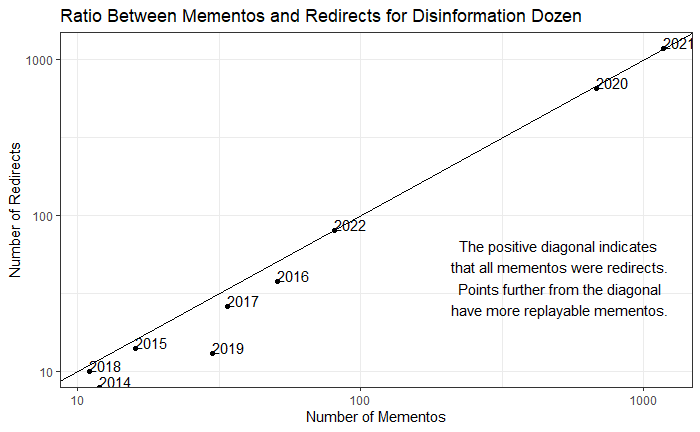}
\vspace*{-3mm}
\caption{Ratio Between Mementos and Redirects for Disinformation Dozen}
\label{fig:disinfo_ratio}
\end{figure}

\begin{figure}[!h]
\centering
\includegraphics[width=0.8\linewidth]{ 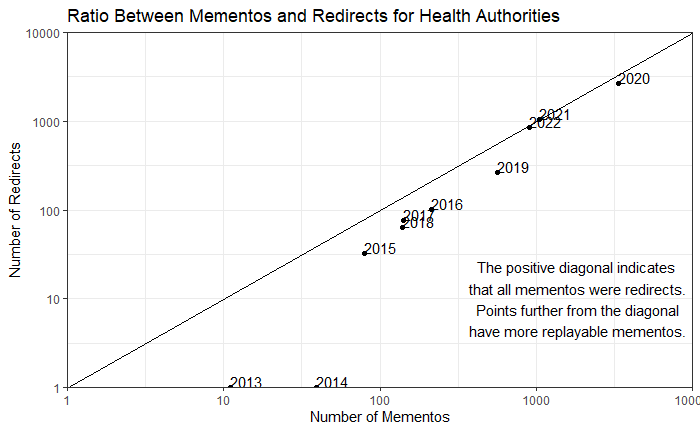}
\vspace*{-3mm}
\caption{Ratio Between Mementos and Redirects for Health Authorities}
\label{fig:health_ratio}
\end{figure}

First of all, we collected the year and HTTP status code of each memento to determine the ratio between the number of total mementos and number of mementos that redirected per year. Figures \ref{fig:disinfo_ratio} and \ref{fig:health_ratio} display this relation for the Disinformation Dozen data set and the Health Authorities data set respectively, with the diagonal line indicating that all mementos for a certain year were redirects.  For example, the points for the years 2021 and 2022 in Figure \ref{fig:disinfo_ratio} appear to be located almost directly on the diagonal line; this positioning illustrates that 1178 mementos redirected out of 1180 total in 2021 (99.83\%), and 80 mementos redirected out of 81 total in 2022 thus far (98.77\%). Similarly, the farther a point is from the diagonal, the more replayable mementos we have for that year. 2014 and 2019 appear to be outliers for the Disinformation Dozen in terms of replayable mementos, with 0 mementos redirecting out of 12 in 2014 (0\%), and 13 mementos redirecting out of 30 in 2019 (43.33\%). Despite the deviancy of 2019, there does appear to be a generalized increase in the number of mementos that are redirects over time. This trend is solidified by the graph for the Health Authorities data set, in which we see three distinct groups over time: 2013 and 2014 with relatively fewer redirects, 2015 through 2019 with approximately half of mementos being redirects, and 2020 through 2022 with almost all mementos being redirects. These results show that we have very few complete user account pages archived for these Instagram accounts| we are losing malicious disinformation campaigns and informative health-related content alike. Finally, it is worth noting that the Health Authorities data set contained 6,472 mementos, which is over three times more than the 2,094 mementos of the Disinformation Dozen data set.

\subsection{What is Happening to the Percentage of Replayable Mementos Over Time?}

\begin{figure}[!!ht]
\centering
\includegraphics[width=0.9\linewidth]{ 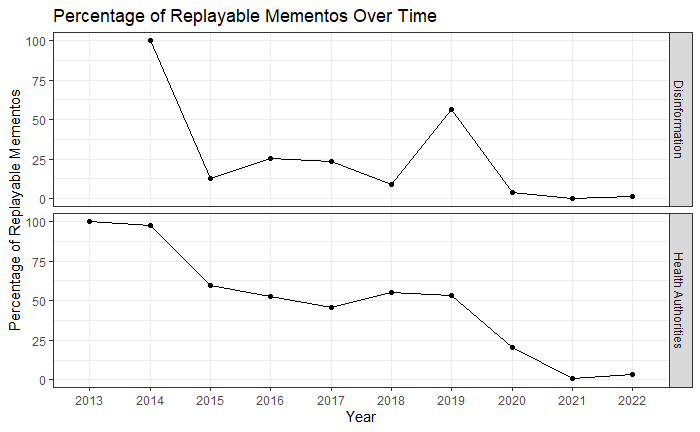}
\vspace*{-3mm}
\caption{Decrease in Percentage of Replayable Mementos Over Time}
\label{fig:percent_replayable}
\end{figure}

In parallel with the increasing number of redirects to the login page over time, both data sets demonstrate a decrease in successful captures of Instagram account pages over time. Furthermore, this decrease is not slight, but extreme; both lines in Figure \ref{fig:percent_replayable} begin with 100\% replayable mementos and end with almost 0\% replayable mementos (1.23\% in 2022 for the Disinformation Dozen and 3.34\% in 2022 for the Health Authorities). Additionally, it can be observed that there are two sharp decreases in each line graph: from 2014 to 2015, and from 2019 to 2020. While we are still solidifying our theory, we suspect that changes to the Instagram user interface during these years made archiving even more difficult for the Wayback Machine crawlers, resulting in more redirects to the login page.
 Overall, these results imply that while the spread of disinformation on Instagram is on the rise, our ability to archive it is on the decline. If we cannot study the divisive content and manipulative strategies of these disinformation actors from recent years, how are we supposed to stop them in the present?  
 \vspace{-2mm}

\vspace{-2mm}
\subsection{How Complete are the Replayable Mementos for Disinformation Dozen and Health Authorities?}

\begin{figure}[h!]
\centering
\includegraphics[width=0.75\linewidth]{ 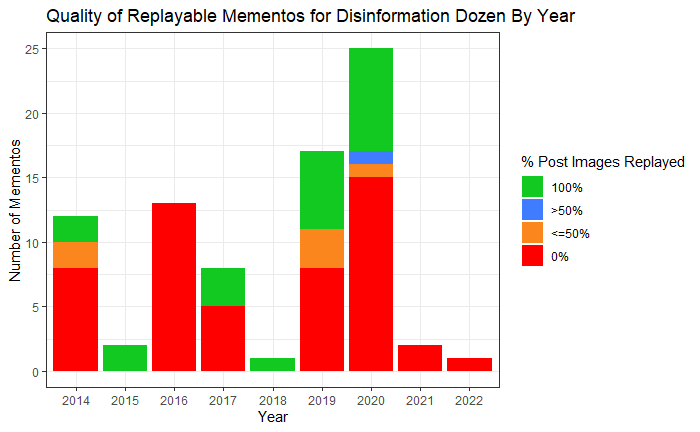}
\vspace*{-4mm}
\caption{Quality of Replayable Mementos for Disinformation Dozen by Year}
\label{fig:bar_disinfo}
\end{figure}

\begin{figure}[h!]
\centering
\includegraphics[width=0.8\linewidth]{ 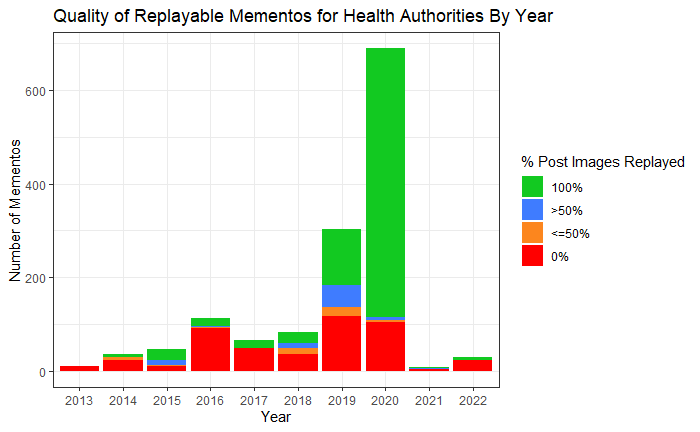}
\vspace*{-4mm}
\caption{Quality of Replayable Mementos for Health Authorities by Year}
\label{fig:bar_health}
\end{figure}

Using the Python script detailed in Section \ref{section:counting_post_images} and the stacked bar chart feature of ggplot2, we were able to illustrate the quality (in terms of complete post images) of the replayable mementos for both groups in Figures \ref{fig:bar_disinfo} and \ref{fig:bar_health}. To clarify, mementos that redirected to the login page were not included in the results, only ``successful" captures of the Instagram account pages. As detailed in the legends, the red portion of each bar indicates that no post images replayed (see Figure \ref{fig:no_loaded_post_images}), the green portion of each bar indicates that every post image replayed (see Figure \ref{fig:complete_loaded_post_images}), and the orange and blue portions indicate that partial post images replayed (see Figure \ref{fig:some_loaded_post_images}), with the former representing under half replaying and the latter representing over half replaying.

Clearly, as demonstrated by the overwhelming amount of red in Figure \ref{fig:bar_disinfo}, a majority of the ``replayable" mementos were not complete, high quality captures of the past. In fact, 64.2\% of replayable mementos for the Disinformation Dozen Instagram profiles had 0 replayable post images and 72.84\%  were incomplete. Additionally, we can observe the significant lack of replayable mementos for the years 2021 and 2022, and furthermore, the poor quality of the few outliers. Finally, it can be noted that 2016 had particularly low quality replayable mementos, and 2019 had a relatively decent number of complete, replayable mementos when compared to the other years.

As for the Health Authorities data set in Figure \ref{fig:bar_health}, one particularly noticeable feature of this graph is the large green portion of the 2020 bar. Because the CDC and the World Health Organization were posting highly critical information about COVID during March of 2020, these account pages were archived at an exceptional rate, resulting in this oversaturation of complete mementos for the year 2020. Regardless of the inflation, we still found that a notable 33.91\% of replayable mementos for the Health Authorities profiles had 0 replayable post images and 43\% were incomplete. Additionally, we see the same distinct lack of mementos for the years 2021 and 2022 that we observed from the Disinformation Dozen data set. Overall, there are clear quality and completeness issues for the replayable mementos, limiting our access to the legitimate content that was posted even further.

\vspace{-5mm}

\section{Conclusion and Future Work}
Concerning the future of this project, a logical subsequent step would be  repeating this analysis on other groups of disinformation actors on Instagram. This study focused on anti-vaxx content creators, but there are other prominent propagators of disinformation that could yield interesting results, such as conspiracy theorists and political figure impersonators. Furthermore, performing the analysis on a control group, such as the mementos for Katy Perry or the Instagram page itself, could be a helpful benchmark for the generalized trends over time. Additionally, as mentioned previously, we plan to investigate when and how the Instagram User Interface has changed over time and to pinpoint which features caused archiving to be more difficult. Finally, we are interested in mapping tagging networks among disinformation actors to identify key players and connections, as well as studying the role of specific hashtags in spreading malicious content over time.

In closing, despite disinformation on Instagram reaching a larger audience and  receiving more engagements than its social media counterparts, it continues to be understudied and underestimated. Due to the strict guidelines that Instagram implements and the consequent banning of disinformation actors, the role of web archiving in preserving and studying the content that was posted cannot be overstated. However, this study took a closer look at what is actually in the archive for Instagram, and found that the presence of a memento does not ensure that the legitimate content was stored. In fact, for the group of anti-vaxx content creators known as the Disinformation Dozen, 96.13\% of the mementos in the Internet Archive's Wayback Machine redirect to the login page. Furthermore, we found that the mementos that do not redirect to the login page may not replay every post image; only 27.16\% of replayable mementos for the Disinformation Dozen replay every post image. Overall, these results signify that only 1.05\% of mementos for the Disinformation Dozen accounts are replayable with complete post images. The loss of around 99\% of malicious content posted is a serious obstacle, because studying the tactics utilitized by disinformation disseminators in the past is the key to preventing the spread of disinformation in the future.  All things considered, finding a more effective way to archive Instagram is an absolute necessity for the web archiving, social media, and disinformation spheres alike.

\section*{Acknowledgements}
This work was supported by NSF CISE REU Site Award \#2149607.

\renewcommand{\refname}{References Cited}
\bibliographystyle{ieeetr}
\bibliography{refs}

\newpage
\appendix
\section*{Appendix}

\begin{table}[h!]
\centering
\begin{tabular}{|c|c|c|c|}
\hline
& Instagram & Facebook & Twitter \\
\hline
Joseph Mercola &  drmercola & doctor.health & mercola \\
Robert F. Kennedy, Jr. & RobertFKennedyJr & rfkjr & RobertKennedyJr \\
Ty and Charlene Bollinger & thetruthaboutvaccinesttav & tycharlene.bollinger & truthaboutbigc \\
Sherri Tenpenny & drtenpenny & VaccineInfo & BusyDrT \\
Rizza Islam & \textunderscore rizzaislam & rizza.infinity & IslamRizza  \\
Rashid Buttar & drbuttar & DrRashidAButtar & DrButtar \\
Erin Elizabeth & healthnutnews & HealthNutNews & unhealthytruth \\
Sayer Ji & greenmedinfo & sayerji & GreenMedInfo \\
Kelly Brogan & kellybroganmd & KellyBroganMD & kellybroganmd \\
Christiane Northrup & drchristianenorthrup & DrChristianeNorthrup & DrChrisNorthrup \\
Ben Tapper & dr.bentapper & DrBenTapper & DrBenTapper1 \\
\hline
\end{tabular}
\caption{Social Media Handles for the Disinformation Dozen}
\label{table:social_media_handles}

\vspace{1.5cm}

\centering
\begin{tabular}{|c|c|c|c|}
\hline
& Instagram & Facebook & Twitter \\
\hline
Internet Archive & 216 & 937 & 871 \\
Archive-It & 2 & 39 & 29 \\
archive.today & 0 & 0 & 20 \\
Library of Congress & 0 & 16 & 24 \\
UK Web Archive & 0 & 2 & 4  \\
Australian Web Archive & 0 & 1 & 0 \\
Portuguese Web Archive & 1 & 0 & 0 \\
\hline
Total & 219 & 995 & 948 \\
\hline
\end{tabular}
\caption{Captures of Joseph Mercola's social media account pages across web archives}
\label{table:joseph_mercola_captures}

\vspace{1.5cm}

\centering
\begin{tabular}{|c|c|c|c|}
\hline
& Instagram & Facebook & Twitter \\
\hline
Internet Archive & 598 & 60 & 193 \\
Archive-It & 0 & 0 & 34 \\
archive.today & 0 & 0 & 55 \\
Library of Congress & 2 & 0 & 1 \\
UK Web Archive & 0 & 0 & 3  \\
Australian Web Archive & 0 & 0 & 0 \\
Portuguese Web Archive & 1 & 0 & 0 \\
\hline
Total & 601 & 60 & 286 \\
\hline
\end{tabular}
\caption{Captures of Robert F. Kennedy Jr.'s social media account pages across web archives}
\label{table:robert_kennedy_captures}
\end{table}

\begin{table}[]
\centering
\begin{tabular}{|c|c|c|c|}
\hline
& Instagram & Facebook & Twitter \\
\hline
Internet Archive & 12 & 4 & 351 \\
Archive-It & 0 & 1 & 290 \\
archive.today & 0 & 0 & 1 \\
Library of Congress & 0 & 0 & 0 \\
UK Web Archive & 0 & 0 & 0  \\
Australian Web Archive & 0 & 0 & 0 \\
Portuguese Web Archive & 0 & 0 & 0 \\
\hline
Total & 12 & 5 & 642 \\
\hline
\end{tabular}
\caption{Captures of the Bollinger's social media account pages across web archives}
\label{table:bollinger_captures}
\end{table}

\begin{table}[]
\centering
\begin{tabular}{|c|c|c|c|}
\hline
& Instagram & Facebook & Twitter \\
\hline
Internet Archive & 362 & 89 & 109 \\
Archive-It & 0 & 13 & 21 \\
archive.today & 0 & 0 & 9 \\
Library of Congress & 0 & 10 & 0 \\
UK Web Archive & 0 & 0 & 0  \\
Australian Web Archive & 0 & 0 & 0 \\
Portuguese Web Archive & 0 & 0 & 0 \\
\hline
Total & 362 & 113 & 140 \\
\hline
\end{tabular}
\caption{Captures of Sherri Tenpenny's social media account pages across web archives}
\label{table:sherri_tenpenny_captures}
\end{table}

\begin{table}[]
\centering
\begin{tabular}{|c|c|c|c|}
\hline
& Instagram & Facebook & Twitter \\
\hline
Internet Archive & 228 & 1 & 17 \\
Archive-It & 0 & 0 & 0 \\
archive.today & 0 & 0 & 7 \\
Library of Congress & 0 & 0 & 0 \\
UK Web Archive & 0 & 0 & 0  \\
Australian Web Archive & 0 & 0 & 0 \\
Portuguese Web Archive & 0 & 0 & 0 \\
\hline
Total & 228 & 1 & 24 \\
\hline
\end{tabular}
\caption{Captures of Rizza Islam's social media account pages across web archives}
\label{table:rizza_islam_captures}
\end{table}

\begin{table}[]
\centering
\begin{tabular}{|c|c|c|c|}
\hline
& Instagram & Facebook & Twitter \\
\hline
Internet Archive & 103 & 27 & 112 \\
Archive-It & 0 & 0 & 18 \\
archive.today & 0 & 0 & 2 \\
Library of Congress & 0 & 0 & 3 \\
UK Web Archive & 0 & 0 & 0  \\
Australian Web Archive & 0 & 0 & 0 \\
Portuguese Web Archive & 0 & 0 & 0 \\
\hline
Total & 103 & 27 & 136 \\
\hline
\end{tabular}
\caption{Captures of Rashid Buttar's social media account pages across web archives}
\label{table:rashid_buttar_captures}
\end{table}

\begin{table}[]
\centering
\begin{tabular}{|c|c|c|c|}
\hline
& Instagram & Facebook & Twitter \\
\hline
Internet Archive & 248 & 146 & 264 \\
Archive-It & 0 & 10 & 110 \\
archive.today & 0 & 0 & 10 \\
Library of Congress & 0 & 1 & 8 \\
UK Web Archive & 0 & 0 & 0  \\
Australian Web Archive & 0 & 0 & 0 \\
Portuguese Web Archive & 0 & 3 & 0 \\
\hline
Total & 248 & 160 & 392 \\
\hline
\end{tabular}
\caption{Captures of Erin Elizabeth's social media account pages across web archives}
\label{table:erin_elizabeth_captures}
\end{table}

\begin{table}[]
\centering
\begin{tabular}{|c|c|c|c|}
\hline
& Instagram & Facebook & Twitter \\
\hline
Internet Archive & 44 & 1058 & 1977 \\
Archive-It & 0 & 15 & 4 \\
archive.today & 0 & 0 & 0 \\
Library of Congress & 0 & 8 & 0 \\
UK Web Archive & 0 & 2 & 0  \\
Australian Web Archive & 0 & 0 & 0 \\
Portuguese Web Archive & 1 & 0 & 0 \\
\hline
Total & 45 & 1083 & 1981 \\
\hline
\end{tabular}
\caption{Captures of Sayer Ji's social media account pages across web archives}
\label{table:sayer_ji_captures}
\end{table}

\begin{table}[]
\centering
\begin{tabular}{|c|c|c|c|}
\hline
& Instagram & Facebook & Twitter \\
\hline
Internet Archive & 22 & 56 & 38 \\
Archive-It & 0 & 0 & 3 \\
archive.today & 0 & 0 & 2 \\
Library of Congress & 0 & 1 & 0 \\
UK Web Archive & 0 & 0 & 0  \\
Australian Web Archive & 0 & 0 & 0 \\
Portuguese Web Archive & 1 & 0 & 0 \\
\hline
Total & 23 & 57 & 43 \\
\hline
\end{tabular}
\caption{Captures of Kelly Brogan's social media account pages across web archives}
\label{table:kelly_brogan_captures}
\end{table}

\begin{table}[]
\centering
\begin{tabular}{|c|c|c|c|}
\hline
& Instagram & Facebook & Twitter \\
\hline
Internet Archive & 91 & 82 & 102 \\
Archive-It & 0 & 11 & 5 \\
archive.today & 0 & 0 & 1 \\
Library of Congress & 0 & 3 & 0 \\
UK Web Archive & 0 & 1 & 0  \\
Australian Web Archive & 0 & 1 & 0 \\
Portuguese Web Archive & 1 & 1 & 1 \\
\hline
Total & 92 & 99 & 109 \\
\hline
\end{tabular}
\caption{Captures of Christiane Northrup's social media account pages across web archives}
\label{table:christiane_northrup_captures}
\end{table}

\begin{table}[]
\centering
\begin{tabular}{|c|c|c|c|}
\hline
& Instagram & Facebook & Twitter \\
\hline
Internet Archive & 87 & 2 & 5 \\
Archive-It & 0 & 0 & 0 \\
archive.today & 0 & 0 & 2 \\
Library of Congress & 0 & 0 & 0 \\
UK Web Archive & 0 & 0 & 0  \\
Australian Web Archive & 0 & 0 & 0 \\
Portuguese Web Archive & 0 & 0 & 0 \\
\hline
Total & 87 & 2 & 7 \\
\hline
\end{tabular}
\caption{Captures of Ben Tapper's social media account pages across web archives}
\label{table:ben_tapper_captures}
\end{table}

\begin{table}[]
\centering
\begin{tabular}{|c|c|c|}
\hline
& Number of Mementos & Percentage of Total Mementos \\
\hline
Identified Redirects & 163 & 73.42\% \\
Warc/Revisits Redirects & 30 & 13.51\% \\
\hline
Total Redirects & 193 & 86.94\% \\
\hline
Identified Successes & 23 & 10.36\% \\
Warc/Revisits Successes & 6 & 2.7\% \\
\hline
Total Successes & 29 & 13.06\% \\
\hline
Total Mementos & 222 & 100.0\% \\
\hline
\end{tabular}
\caption{Redirects vs. successes for instagram.com/drmercola}
\label{table:joseph_mercola_redirects}
\end{table}

\begin{table}[h]
\centering
\begin{tabular}{|c|c|c|}
\hline
& Number of Mementos & Percentage of Total Mementos \\
\hline
Identified Redirects & 574 & 95.99\% \\
Warc/Revisits to Redirects & 18 & 3.01\% \\
\hline
Total Redirects & 592 & 99.0\% \\
\hline
Identified Successes & 6 & 1.0\% \\
Warc/Revisits to Successes & 0 & 0.0\% \\
\hline
Total Successes & 6 & 1.0\% \\
\hline
Total Mementos & 598 & 100.0\% \\
\hline
\end{tabular}
\caption{Redirects vs. successes for instagram.com/RobertFKennedyJr}
\label{table:robert_kennedy_redirects}
\end{table}

\begin{table}[h]
\centering
\begin{tabular}{|c|c|c|}
\hline
& Number of Mementos & Percentage of Total Mementos \\
\hline
Identified Redirects & 10 & 83.33\% \\
Warc/Revisits Redirects & 2 & 16.67\% \\
\hline
Total Redirects & 12 & 100.0\% \\
\hline
Identified Successes & 0 & 0.0\% \\
Warc/Revisits Successes & 0 & 0.0\% \\
\hline
Total Successes & 0 & 0.0\% \\
\hline
Total Mementos & 12 & 100.0\% \\
\hline
\end{tabular}
\caption{Redirects vs. successes for instagram.com/thetruthaboutvaccinesttav}
\label{table:bollinger_redirects}
\end{table}

\begin{table}[]
\centering
\begin{tabular}{|c|c|c|}
\hline
& Number of Mementos & Percentage of Total Mementos \\
\hline
Identified Redirects & 350 & 95.63\% \\
Warc/Revisits Redirects & 11 & 3.01\% \\
\hline
Total Redirects & 361 & 98.63\% \\
\hline
Identified Successes & 5 & 1.37\% \\
Warc/Revisits Successes & 0 & 0.0\% \\
\hline
Total Successes & 5 & 1.37\% \\
\hline
Total Mementos & 366 & 100.0\% \\
\hline
\end{tabular}
\caption{Redirects vs. successes for instagram.com/drtenpenny}
\label{table:sherri_tenpenny_redirects}
\end{table}

\begin{table}[]
\centering
\begin{tabular}{|c|c|c|}
\hline
& Number of Mementos & Percentage of Total Mementos \\
\hline
Identified Redirects & 228 & 99.13\% \\
Warc/Revisits Redirects & 0 & 0.0\% \\
\hline
Total Redirects & 228 & 99.13\% \\
\hline
Identified Successes & 2 & 0.87\% \\
Warc/Revisits Successes & 0 & 0.0\% \\
\hline
Total Successes & 2 & 0.87\% \\
\hline
Total Mementos & 230 & 100.0\% \\
\hline
\end{tabular}
\caption{Redirects vs. successes for instagram.com/\_rizzaislam}
\label{table:rizza_islam_redirects}
\end{table}

\begin{table}[]
\centering
\begin{tabular}{|c|c|c|}
\hline
& Number of Mementos & Percentage of Total Mementos \\
\hline
Identified Redirects & 100 & 95.24\% \\
Warc/Revisits Redirects & 1 & 0.95\% \\
\hline
Total Redirects & 101 & 96.19\% \\
\hline
Identified Successes & 4 & 3.81\% \\
Warc/Revisits Successes & 0 & 0.0\% \\
\hline
Total Successes & 4 & 3.81\% \\
\hline
Total Mementos & 105 & 100.0\% \\
\hline
\end{tabular}
\caption{Redirects vs. successes for instagram.com/drbuttar}
\label{table:rashid_buttar_redirects}
\end{table}

\begin{table}[]
\centering
\begin{tabular}{|c|c|c|}
\hline
& Number of Mementos & Percentage of Total Mementos \\
\hline
Identified Redirects & 264 & 90.72\% \\
Warc/Revisits Redirects & 3 & 1.03\% \\
\hline
Total Redirects & 267 & 91.75\% \\
\hline
Identified Successes & 23 & 7.9\% \\
Warc/Revisits Successes & 1 & 0.34\% \\
\hline
Total Successes & 24 & 8.25\% \\
\hline
Total Mementos & 291 & 100.0\% \\
\hline
\end{tabular}
\caption{Redirects vs. successes for instagram.com/healthnutnews}
\label{table:erin_elizabeth_redirects}
\end{table}

\begin{table}[]
\centering
\begin{tabular}{|c|c|c|}
\hline
& Number of Mementos & Percentage of Total Mementos \\
\hline
Identified Redirects & 34 & 77.27\% \\
Warc/Revisits Redirects & 4 & 9.09\% \\
\hline
Total Redirects & 38 & 86.36\% \\
\hline
Identified Successes & 6 & 13.64\% \\
Warc/Revisits Successes & 0 & 0.0\% \\
\hline
Total Successes & 6 & 13.64\% \\
\hline
Total Mementos & 44 & 100.0\% \\
\hline
\end{tabular}
\caption{Redirects vs. successes for instagram.com/greenmedinfo}
\label{table:sayer_ji_redirects}
\end{table}

\begin{table}[]
\centering
\begin{tabular}{|c|c|c|}
\hline
& Number of Mementos & Percentage of Total Mementos \\
\hline
Identified Redirects & 13 & 56.52\% \\
Warc/Revisits Redirects & 8 & 34.78\% \\
\hline
Total Redirects & 21 & 91.3\% \\
\hline
Identified Successes & 2 & 8.7\% \\
Warc/Revisits Successes & 0 & 0.0\% \\
\hline
Total Successes & 2 & 8.7\% \\
\hline
Total Mementos & 23 & 100.0\% \\
\hline
\end{tabular}
\caption{Redirects vs. successes for instagram.com/kellybroganmd}
\label{table:kelly_brogan_redirects}
\end{table}

\begin{table}[]
\centering
\begin{tabular}{|c|c|c|}
\hline
& Number of Mementos & Percentage of Total Mementos \\
\hline
Identified Redirects & 90 & 92.78\% \\
Warc/Revisits Redirects & 5 & 5.15\% \\
\hline
Total Redirects & 95 & 97.94\% \\
\hline
Identified Successes & 2 & 2.06\% \\
Warc/Revisits Successes & 0 & 0.0\% \\
\hline
Total Successes & 2 & 2.06\% \\
\hline
Total Mementos & 97 & 100.0\% \\
\hline
\end{tabular}
\caption{Redirects vs. successes for instagram.com/drchristianenorthrup}
\label{table:christiane_northrup_redirects}
\end{table}

\begin{table}[]
\centering
\begin{tabular}{|c|c|c|}
\hline
& Number of Mementos & Percentage of Total Mementos \\
\hline
Identified Redirects & 104 & 100.0\% \\
Warc/Revisits Redirects & 0 & 0.0\% \\
\hline
Total Redirects & 104 & 100.0\% \\
\hline
Identified Successes & 0 & 0.0\% \\
Warc/Revisits Successes & 0 & 0.0\% \\
\hline
Total Successes & 0 & 0.0\% \\
\hline
Total Mementos & 104 & 100.0\% \\
\hline
\end{tabular}
\caption{Redirects vs. successes for instagram.com/dr.bentapper}
\label{table:ben_tapper_redirects}
\end{table}

\begin{table}[t!]
\centering
\begin{tabular}{|c|c|}
\hline
& Instagram  \\
\hline
BBC News &  bbcnews\\
UNICEF & unicef \\
The CDC & cdcgov \\
WHO & who \\
Bill Gates & thisisbillgates  \\
UK Government & ukgovofficial \\
NHS & nhs \\
Gates Foundation & gatesfoundation  \\
London School of Hygiene and Tropical Medicine & lshtm \\
\hline
\end{tabular}
\caption{Instagram Handles for the Health Authorities}
\label{table:instagram_handles}
\end{table}

\newpage

\end{document}